# Signal to noise ratio of phase sensing telescope interferometers

**François Hénault**
*Observatoire de la Côte d'Azur, Laboratoire Gemini, UMR 6203, CNRS*
*Avenue Nicolas Copernic, 06130 Grasse, France*

Abstract: This paper is the third part of a trilogy dealing with the principles, performance and limitations of what I named "Telescope-Interferometers" (TIs). The basic idea consists in transforming one telescope into a Wavefront Error (WFE) sensing device. This can be achieved in two different ways, namely the off axis and phase-shifting TIs. In both cases the Point-Spread Function (PSF) measured in the focal plane of the telescope carries information about the transmitted WFE, which is retrieved by fast and simple algorithms suitable to an Adaptive Optics (AO) regime. Herein are evaluated the uncertainties of both types of TIs, in terms of noise and systematic errors. Numerical models are developed in order to establish the dependence of driving parameters such as useful spectral range, angular size of the observed star, or detector noise on the total WFE measurement error. The latter is found particularly sensitive to photon noise, which rapidly governs the achieved accuracy for telescope diameters higher than 10 m. We study a few practical examples, showing that TI method is applicable to AO systems on telescope diameters ranging from 10 to 50 m, depending on seeing conditions and magnitude of the observed stars. We also discuss the case of a space-borne coronagraph where TI technique provides high sampling of the input WFE map.

*OCIS codes: 010.7350, 070.2580, 100.5070, 110.5100, 110.6770*

## 1. Introduction

Recently, a new approach was suggested for WFE sensing [1-2], combining the advantages of direct phase measurement performed at the telescope focal plane, on one hand, and the possibility to work under an Adaptive Optics (AO) regime, since only simple numerical algorithms compatible with real-time operation are employed, on the other hand. Indeed, the basic idea consists in transforming the telescope itself into a phase sensing apparatus by creating sets of interference fringes into the focal plane. Practically, this is realized by adding one reference arm at the pupil plane of the telescope – a modification requiring subsequent additional opto-mechanical hardware. Information about Wavefront Error (WFE) is then extracted from the measured signal by means of an inverse Fourier transform. The device, which I called "Telescope-Interferometer" (TI), belongs to a new generation of Wavefront Sensors (WFS) that are installed at the image plane of the telescope, rather than at the pupil plane. Such kind of WFS already exists: let us mention for examples works from Angel [3] and Labeyrie [4], whose proposed designs are based on a Mach-Zehnder interferometer, eventually using holographic techniques.

AO principle was first proposed by Babcock [5] in 1953, and encountered continually growing success after a few decades. Today the largest observatories on Earth are all equipped with this technology that demonstrates outstanding capacities to pass beyond the seeing limit and reveal unsuspected details about numerous types of sky objects, particularly in the infrared region of the electromagnetic spectrum. But adaptive optics is continually faced to new challenges, such as covering low-wavelength spectral domain. In addition, AO has to cope with the increasing size of future Extremely Large Telescopes (ELTs), with diameters ranging from 30 to 50 meters. For such facilities, it is expected that the primary mirror will be made of an array of smaller reflecting segments, like the Keck, GranTeCan or JWST (James Webb Space Telescope) already are. In that case, one of the most critical problems becomes to adjust (or co-phase) the individual pistons of the segments in order to approximate the continuous theoretical surface of the primary mirror within accuracies typically better than one tenth of wavelength. Looking deeper into the future, the imaging hyper-telescope proposed by Labeyrie [6] will also impose to develop robust co-phasing capacities.

The measurement of the WFE emerging from a telescope can be carried out in several different ways. In the field of AO, the most common method is to sense WFE by means of a pupil plane WFS, such as Shack-Hartmann [7], curvature [8], pyramidal [9] or optical differentiation sensors [10]. Generally speaking however, these devices are not suitable for co-phasing mirror segments, because their basic principle consists in measuring phase slopes or curvatures and then retrieving WFE using digital procedures. Hence they should not recognize piston errors. To overcome this difficulty, two different approaches were followed.

1) Based on legacy of cophasing experiments at the Keck segmented primary mirror [11-12], several active optics systems were developed during recent years. They consist in upgraded versions of some of the previous AO WFS, having the capacity to discriminate piston errors by means of enhanced hardware and algorithms. Their common principle is to detect local slope or curvature breaks of the measured WFEs. Hence advanced Shack-Hartmann, curvature [13] and pyramidal [14] WFS were successfully built and tested, as well as a modified Mach-Zehnder interferometer [15] located in the telescope focal plane. It must be noted that European Southern Obervatory (ESO) recently developed an Active Phasing Experiment (APE) in order to compare the performance of those sensors in laboratory conditions and on-sky [16].

2) Another way is to employ image plane restoration techniques described by Fienup [17] or Gonsalves [18], respectively phase retrieval and phase diversity, which were subject to numerous software developments over the years.

All the previous methods, however, are not well matched to AO operation because they usually require significant post-processing times. The "ideal" wavefront sensor should indeed combine the ability to perform direct WFE measurements in quasi real-time. Thus the main purpose of this paper is to evaluate the real capacity of Telescope-Interferometers to fulfill this objective. The general principles of TIs are summarized in section 2. Then a comprehensive error analysis of both types of TIs is provided in section 3, including random noise as well as systematic errors (or bias). Typical TI applications and their range of validity are discussed in section 4, before general conclusions are given in section 5.

## 2. About Telescope-Interferometers

In this section are summarized the basic principles of Telescope-Interferometers as they are presented in Refs. [1-2]. Actually two different types of TI were proposed, which are the off-axis TI [1] and the phase-shifting TI [2]. Although both versions look quite different at first glance, they are essentially based on the same idea. In the case of an off-axis TI, a decentred, reference telescope is added aside the main pupil (where the WFE is to be measured), at a distance B from the optical axis (see Figure 1). Both apertures are focusing light at the same point O', that is the telescope focus. There a spatially modulated Point Spread Function (PSF) is generated in the focal plane and measured by means of a detector array. It can be shown that the Optical Transfer Function (OTF) of the system – computed via inverse Fourier transform – provides quantitative information about the telescope WFE. In its phase-shifting version (see Figure 2), the reference pupil is located inside the main aperture rim. It is a mobile part introducing calibrated phase steps $\phi$ when shifted along the optical axis. Then several PSFs are measured and Fourier transformed, before their OTFs are linearly combined in order to retrieve the original WFE. In both cases, one fundamental prerequisite is that the diameter of the reference pupil must be significantly smaller than the diameter of the telescope to be measured, i.e. by one order of magnitude at least.

     Detailed theories of off-axis and phase-shifting TIs are provided in Refs. [1-2] respectively. For the sake of understanding however, some basic steps are reminded here after. Let us denote $A_R$ the amplitude of the incident electric field and $B_R(x,y)$ the two-dimensional transmission function of the main pupil, uniformly equal to one inside of its contours, and to zero anywhere else – note that this function is not generally circular, and R stands for the clear aperture of the telescope. The incoming light is assumed to be monochromatic at a wavelength $\lambda$. We first consider the case of an off-axis TI.

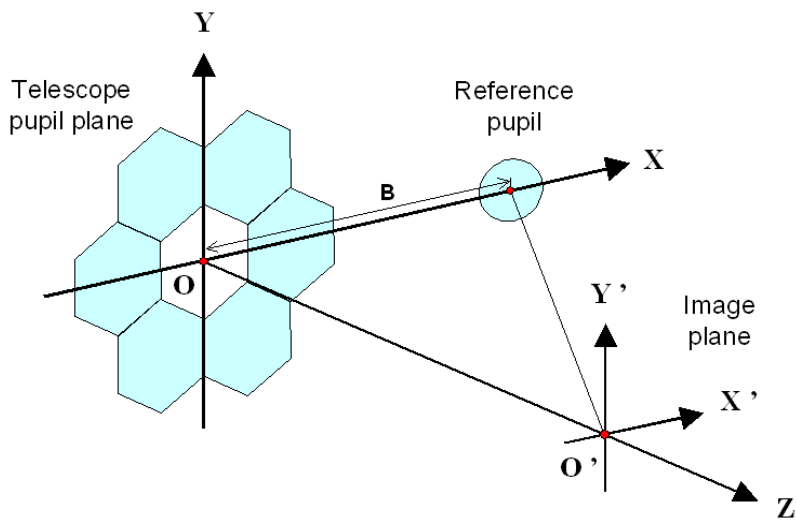

**Figure 1: Schematic drawing of an off-axis Telescope-Interferometer**

## 2.1 Off-axis TI

The wave emerging from both main and reference apertures of the off-axis TI can be written in the pupil plane OXY:

$$A_P(x,y) = A_R B_R(x,y) \exp[ik\Delta(x,y)] + A_r B_r(x-B,y) \quad (1)$$

where $k = 2\pi/\lambda$, $A_r$ is the amplitude of light on the reference pupil, $B_r(x,y)$ its transmission map equal to the "top-hat" function of radius r, and B the distance separating both aperture centers along X-axis. The wave diffracted in the O'X'Y' plane is then obtained via Fourier transforming of $A_P(x,y)$:

$$A'_P(x',y') = FT[A_P(x,y)] = \iint_{x,y} A_P(x,y) \exp[-i 2\pi(ux+vy)] dx\, dy$$

$$= A_R FT\{B_R(x,y) \exp[ik\Delta(x,y)]\} + A_r FT\{B_r(x-B,y)\} \quad (2)$$

with $u = x'/\lambda F$ and $v = y'/\lambda F$, and F the telescope focal length. Then the Point Spread Function PSF(x',y') in the focal plane of the telescope-interferometer is equal to the square modulus of $A_P'(x',y')$:

$$PSF(x',y') = |A'_P(x',y')|^2 \quad (3)$$

Phase retrieval procedure now consists in acquiring this PSF, either using CCD detector arrays or photon-counting cameras, and to compute digitally the OTF of the TI by means of an inverse Fourier transform, i.e.

$$OTF(x,y) = FT^{-1}[PSF(x',y')] \quad (4)$$

It has been demonstrated that, taking into account the effective areas of the main and reference pupils respectively denoted $S_R$ and $S_r$, and using some classical properties of Fourier transforms and convolution products, a rigorous expression of the OTF as a function of (x,y) coordinates can be defined as [1]:

$$OTF(x,y) = A_R^2 S_R^2 \left\{ \frac{1}{S_R} O_R(x,y) + C^2 \frac{1}{S_r} O_r(x,y) + C \frac{1}{S_R} \left(B_R(x,y) \exp[ik\Delta(x,y)]\right) \otimes B_r(x+B,y)/S_r \right.$$

$$\left. + C \frac{1}{S_R} \left(B_R(-x,-y) \exp[-ik\Delta(-x,-y)]\right) \otimes B_r(x-B,y)/S_r \right\} \quad (5)$$

where ⊗ denotes a convolution product, and C is the "contrast ratio" of the TI, which is a key parameter with regard to its capacity and performance – it must be noticed that $A_r = A_R$ most of the times, so the TI contrast often reduces to the ratio between the geometrical surfaces of reference and main pupils.

$$C = \frac{A_r}{A_R} \times \frac{S_r}{S_R} \qquad (6)$$

The OTF in Eq. (5) is composed of four terms. We recognize in the first two ones the Modulation Transfer Function (MTF) of the main and reference pupils, each considered individually. But actually the most important term is the third, because it provides evidence of a convolution relationship between the reference pupil and the WFE to be measured. Furthermore, it can easily be extracted from the global OTF in Eq. (5) since it is de-centered of –B along the X-axis. Hence if the following condition is respected:

$$B \geq 3R + r, \qquad (7)$$

there can be no crosstalk with the other OTF terms, and the third one can be written as:

$$A_R^2 S_R C \left( B_R(x,y) \exp[ik\Delta(x,y)] \right) \otimes B_r(x,y)/S_r = OTF(x-B,y) B_{R+r}(x,y) \qquad (8)$$

Equation (8) could be resolved by a deconvolution process, but at the price of extensive computation times, making the method inappropriate for quasi real-time measurements and thus adaptive optics. Instead, it can be assumed that the function $B_r(x,y)/S_r$ tends towards a Dirac distribution, an approximation all the more valid as the ratio $S_r/S_R$ is small with respect to unity. Then the WFE retrieval relationship becomes:

$$A_R^2 S_R C B_R(x,y) \exp[ik\Delta(x,y)] \approx OTF(x-B,y) B_{R+r}(x,y) \qquad (9)$$

and the searched WFE is directly proportional to the phase of the computed OTF. It must be emphasized that the previous "Dirac approximation" is indeed the heart of the TI measurement method. As a consequence, spatial information available on the optical surface of the main telescope is constrained by the size of the reference pupil, and the spatial resolution of a TI can be estimated around $(S_R/S_r)^{1/2}$, even if spatial sampling on the pupil largely exceeds this limit, since sampling is related to the number of pixels of the camera used for PSF data acquisitions. Also, it can be felt intuitively that, if spatial resolutions are increased by defining smaller reference pupils, this will be detrimental to the TI contrast ratio, and then to the measurement noise. The point is of concern for all types of TIs, and is addressed in sections 3 and 4.

Another consequence of Eq. (9) is that the capture range of $\Delta(x,y)$ is limited to one wavelength at most, and phase unwrapping procedures must be implemented in order to reconstruct the original WFE. Practically, this can be achieved by means of classical unwrapping algorithms such as described in Ref. [19], or by performing several OTF calculations at different wavelengths as suggested by Löfdahl and Eriksson [20].

## 2.2 Phase-shifting TI

Let us now briefly summarize the case of the phase-shifting TI, since the general approach merely follows the same steps. The reference pupil is then located at the center of the primary mirror of the main telescope – see Figure 2. It must be noticed that this is a general illustration and that drawings showing real optical schemes and how light crosses the obscured part of pupil are provided in Ref. [2]. Let us assume that this reference pupil can be moved along the Z optical axis of known quantities corresponding to phase shifts $\phi$. The amplitude of the wave in the pupil plane OXY now writes:

$$A_P(x,y) = A_R B_R(x,y) \exp[ik\Delta(x,y)] + A_r B_r(x,y) \exp[i\phi] \qquad (10)$$

and the amplitude in the image plane O'X'Y' will be:

$$A'_P(x',y') = A_R FT\{B_R(x,y)\exp[ik\Delta(x,y)]\} + A_r \exp[i\phi] FT\{B_r(x,y)\} \qquad (11)$$

As in the previous section, this brings about a four-terms expression of the OTF, and the third one can be isolated by acquiring four PSFs with different values of the varying pupil shift $\phi$ (i.e. $\phi = 0$, $\pi/2$, $\pi$ and $3\pi/2$). The afore mentioned "Dirac approximation" is also employed, finally leading to an estimate of the incident wave:

$$A_R{}^2 S_R C B_R(x,y)\exp[ik\Delta(x,y)] \approx [OTF_0(x,y) + iOTF_{\pi/2}(x,y) - OTF_\pi(x,y) - iOTF_{3\pi/2}(x,y)]/4 \qquad (12)$$

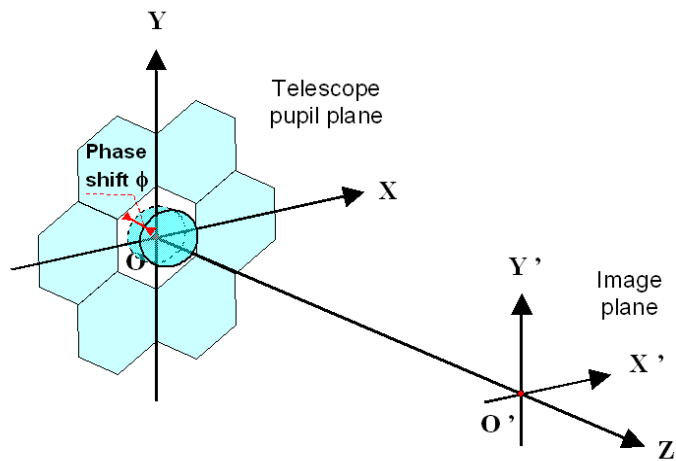

**Figure 2: Schematic drawing of a phase-shifting TI**

To conclude, the principles of two phase sensing telescope-interferometers (respectively the off-axis and phase-shifting TIs) were briefly summarized in this section. The here above basic relationships will prove very helpful in the following section, where we seek to evaluate measurement accuracies of the techniques, and to assess the actual capacities of both TI families.

## 3. Analysis of WFE restoration errors

An extensive, in-depth analysis of all types of systematic errors and random noises that can affect TIs performance would probably require the writing of a full book, and the completion of a subsequent error budget may be quite laborious, even if that work will need to be carried out if a TI has to be built someday. At the moment, I chose to select a few major error sources, trying to estimate their consequences, and determine if they correspond to theoretical barriers for practical implementation of TIs. Additionally, the knowledge of such errors items allows to define the best application ranges for those phase sensing devices, as discussed in section 4. Herein are distinguished two kinds of measurement uncertainties that are bias or systematic errors, on one hand, and random error noises, on the other hand. Bias errors may result from physical limitations such as intrinsic error due to the "Dirac approximation" in Eq. (9) and (12), useful spectral bandwidth, angular size or magnitude of the target star, and differential WFEs between reference and main pupils – the latter point, however, was already discussed in Ref. [1]. On their side, random errors originate from the used detection system, i.e. the CCD camera, but may also include atmospheric turbulence (which will be discussed in section 4.2) and accuracy of piston displacements of the reference mirror in the case of a phase-shift TI. To be exhaustive we should also mention telescope aberrations, polishing defects of mirror segments and eventual coupled terms between all previous error sources. But the effects of photon noise are of particular interest here, since the useful signal is proportional to the contrast ratio C, a quantity that is significantly lower than unity for both types of TIs. Hence the influence of Signal-to-Noise Ratio (SNR) is expected to be the most critical issue for what concerns measurement accuracy. Fortunately, a simplified expression can be derived in the presence of random detection errors.

### *3.1 Random detection errors*

Typical random detection errors are photon noise, Read-Out Noise (RON), dark current of the detector array, and the way they depend on parameters such as integration time or working temperature of the CCD chip. Those noises add up to the theoretical signal delivered by the TI, which is the complex quantity $A_R^2\ S_R\ C\ B_R(x,y)\ \exp[i\ k\ \Delta(x,y)]$ defined by Eqs. (9) and (12) at a given point of pupilar coordinates (x,y). We first rewrite $B_R(x,y)$ as:

$$B_R(x,y) = \eta\{1 + \delta B_R(x,y)\} \quad (13)$$

where $\eta$ is the global radiometric efficiency of the TI from sky photons to output detected electrons, including atmospheric transmittance, telescope optical losses on reflective and refractive surfaces and the CCD detector spectral response curve. $\delta B_R(x,y)$ stands for small variations of the TI transmission, either generated by detector noise or atmospheric scintillation – here it must be emphasized that through relationships (9) or (12) TIs present an additional capacity for scintillation measurements, with a spatial resolution limited by the size of the reference pupil. Let us denote the noise term resulting from the inverse Fourier transform in Eq. (4) as a complex number of modulus $\sigma_\Delta$ and phase $\psi_\Delta$, such as represented in Figure 3. This term takes into account $\delta B_R(x,y)$ as well as small deviations of the telescope WFE, noted $\delta\Delta(x,y)$. Hence we write:

$$\eta A_R^2 S_R C B_R(x,y) \exp[ik\Delta(x,y)] + \sigma_\Delta \exp[i\psi_\Delta] = \eta A_R^2 S_R C \{1+\delta B_R(x,y)\} \exp[ik(\Delta(x,y)+\delta\Delta(x,y))] \quad (14)$$

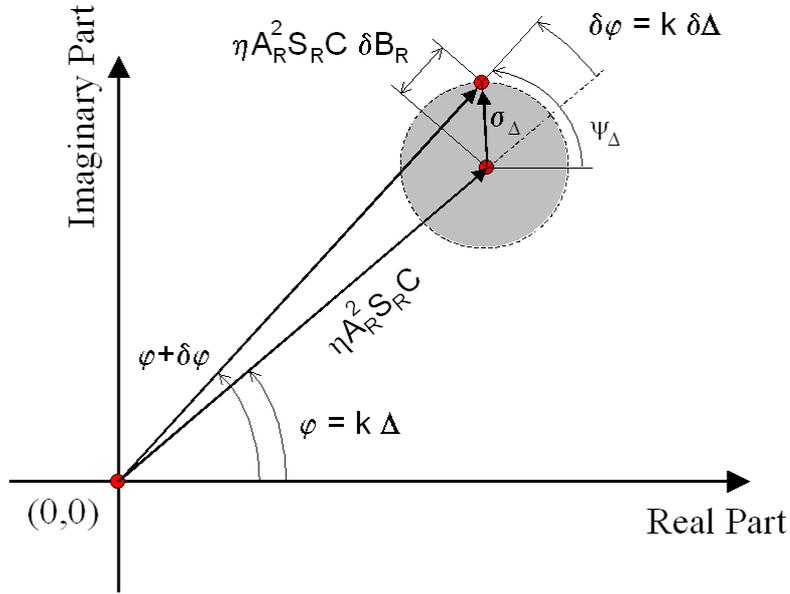

**Figure 3: Signal and noise addition in complex plane**

Let us now assume that $\delta B_R(x,y)$ and $\delta\Delta(x,y)$ are small terms with regard to 1 and $\Delta(x,y)$ respectively. Then a first-order approximation of Eq. (14) reduces to:

$$\sigma_\Delta \exp[i\psi_\Delta] = \eta A_R^2 S_R C [\delta B_R(x,y)+ik\delta\Delta(x,y)] \times \exp[ik\Delta(x,y)] \quad (15)$$

When only modules of complex quantities are considered, noise expression becomes:

$$\sigma_\Delta = \eta A_R^2 S_R C [\delta B_R^2(x,y)+k^2\delta\Delta^2(x,y)]^{1/2} \quad (16)$$

The previous relationship is the basis for the estimation of TIs measurement noises. Assuming that the term $\delta B_R(x,y)$ is equal to zero, we obtain a worst-case, maximal bound for $\delta\Delta(x,y)$:

$$k\,\delta\Delta(x,y) \leq \sigma_\Delta / \eta A_R^2 S_R C \qquad (17)$$

Moreover, we recognize the product $A_R^2 S_R$ as the total flux impinging on the main telescope pupil, and the ratio $\eta A_R^2 S_R / \sigma_\Delta$ as the effective SNR of the telescope-interferometer, since noise modulus is invariant via the inverse Fourier transform in relation (4), as stated by Parseval's theorem. Thus Eq. (17) can be rewritten as one of the simplest expressions ever obtained when evaluating WFS measurement errors (see for example Ref. [21], section 5):

$$k\,\delta\Delta(x,y) \leq 1/(SNR \times C) \qquad (18)$$

The SNR is computed according to the usual "CCD equation" which has the form [21-22]:

$$SNR = \eta P \tau / [\eta P \tau + n_{Pix}(d\tau + r^2)]^{1/2} \qquad (19)$$

where P is the total flux expressed in photons/second and $\tau$ is the integration time on the CCD array. Here the first term under the square root at the denominator stands for photon noise. Other parameters appearing in the formula are the number of pixel rows $n_{Pix}$, the dark current noise d expressed in electrons/photon/second/pixel, and r the RON of the detector array in terms of electrons/photon/pixel. The rigorous expression of P classically is the product of the brightness $B_P(\lambda)$ of a blackbody integrated over a spectral bandwidth $\Delta\lambda$, with the geometrical etendue of the beam collected by the main telescope, i.e.

$$P = S_R \pi \varepsilon^2 \int_{\Delta\lambda} B_P(\lambda) d\lambda \qquad (20)$$

with $\varepsilon$ being the angular radius of the target star. The here above set of equations (18-20) applies to both types of TIs, enabling to compute their measurement accuracies as functions of various types of random errors, including in particular photon noise. Relation (18) also shows that the achieved precision is proportional to the inverse of the contrast ratio C, itself related to TI spatial resolution. Thus the concluding remark in section 2.1 gets its confirmation here: high TI spatial resolutions tend to increase noises and consequently to deteriorate the global measurement accuracy. This implies that trade-offs between those both physical characteristics need to be carried out for optimal WFE sensing.

### 3.2 Systematic or bias errors

The question of systematic, or "bias" errors has already been evoked in Ref. [1], where a preliminary, non-exhaustive list of error sources was identified and discussed. However those topics need to be revisited herein, in order to obtain quantitative estimations and to define acceptable limits from which specific requirements might be put on the observed sky-objects. Let us first remind three major quoted systematic errors.

1) Intrinsic error due to the "Dirac approximation". It was already seen in section 2 that Eqs. (9) and (12) – respectively for the off-axis and phase-shifting TIs – result from the assumption that the area of the reference pupil is sensibly smaller than the main telescope aperture, thus allowing to replace the reference pupil transmission function $B_r(x,y)/S_r$ with the theoretical Dirac distribution. Eliminating this "Dirac approximation" would clearly be the best way to enhance accuracy near the rim and central obstruction of the main aperture. This can only be achieved by means of sophisticated deconvolution algorithms, since rather simple processes such as direct deconvolution were already tested, and showed no noticeable improvements with respect to the hypothesis of a point-like reference pupil. Hence when using the same first-order approximation than in Eq. (15), the induced measurement error $\delta\Delta(x,y)$ can be written mathematically:

$$k\delta\Delta(x,y) = -i\{(\exp[ik\Delta(x,y)]\otimes B_r(x,y)/S_r) - \exp[ik\Delta(x,y)]\}/\exp[ik\Delta(x,y)] \quad (21)$$

Unfortunately the latter expression is inappropriate for additional analytic developments, thus $\delta\Delta(x,y)$ needs to be evaluated digitally. Numerical simulations were performed in Ref. [1], showing that achievable accuracy depends on the WFE $\Delta(x,y)$ to be measured, on one hand, and that the main errors are encountered near the main pupil rim, on the other hand. Among different types of WFEs, the case of segmented mirrors suffering from varied piston errors proved to be one of the most challenging, and is obviously of major concern here. Therefore updated numerical simulations are provided in section 3.3.

2) Effect of spectral bandwidth. For both types of TIs, PSF calculations and phase retrieval algorithms defined in sections 2.1 and 2.2 are fundamentally monochromatic. But even when using spectral filters, such theoretical assumption cannot be achieved in practice. Since the width of a PSF is classically proportional to the real wavelength $\lambda$, the different PSFs will add up in the focal plane, and the expression of the actually retrieved OTF – as defined by Eq. (4) – becomes:

$$OTF(x,y) = FT^{-1}\left[\int_{\Delta\lambda} PSF_\lambda(x',y')d\lambda\right] \quad (22)$$

where the subscript $\lambda$ highlights dependence of the PSF with respect to wavelength. Then $OTF(x,y)$ is itself affected by a bias error that will propagate in data handling until the final WFE expressions $\Delta(x,y)$ in Eqs. (9) and (12). Here again, no analytical development seems easily feasible and the systematic error has to be computed numerically.

3) Apparent size of the target star. Here is considered the case when the target star is partially or totally resolved by the TI. As in previous case, Eq. (4) defining the resulting OTF must be modified since the PSF is now convoluted by the apparent size of the star (assumed to be circular with angular radius $\varepsilon$), multiplied by the focal length F of the main telescope.

$$OTF(x,y) = FT^{-1}\left[PSF(x',y') \otimes B_{F\epsilon}(x',y')\right] \qquad (23)$$

Therefore an additional bias error term related to the dimensions of the observed astronomical source is generated. It must be noticed that an analytic, rigorous expression of OTF(x,y) can be derived from Eq. (23), since the inverse Fourier transform of $B_{F\epsilon}(x',y')$ is proportional to the Airy function $2J_1(\rho)/\rho$ where $J_1$ is the type-J Bessel function at the first order, acting as a multiplying factor in the WFE retrieval formulae (9) and (12). But from a practical point of view, this error term cannot be easily separated from the intrinsic error defined by Eq. (21). Hence a complete numerical estimation of all bias errors was finally preferred.

After having identified and discussed three major systematic error terms of the TI method and given their theoretical formulae in Eqs. (21-23), we arrive at the conclusion that their order of magnitude should better be estimated numerically with help from a computer model. This is the scope of the next section.

## 3.3 Computer Modeling

In fact, two TIs numerical models have already been elaborated under IDL language, and some significant results were released in Refs. [1-2], respectively for what concerns the off-axis and phase shifting versions. However the general architecture of used computer programs had to be deeply modified in order to support the cases of polychromatic, extended sky-objects mentioned in the previous section. The TI computer model is now split into three major modules (see Figure 4).

1) A FORTRAN ray-tracing software is in charge of computing the PSFs measured at the TIs focal plane. It has the ability of modeling different apertures of various shape and sizes, taking into account the real geometry and alignment errors – in both piston and tip-tilt – of segmented mirrors. The resulting wavefront error is stored in an external "WFE file" for upcoming comparison with WFEs retrieved by TIs. For each given wavelength and angular direction, the program computes the PSF according to Fraunhofer's approximation, i.e. via a direct Fourier transform. The PSFs are then added together while several loops are executed over the useful spectral bandwidth and angular size of the sky-object for different wavelengths and spatial directions. Finally, the program samples the obtained irradiances at the detector surface, eventually adding statistical noise errors – including photon noise – and stores the computed data in an external "PSF file". For a phase-shifting TI, this program is run four times, with each different value of the reference phase shift $\phi$ (see section 2.2).
2) The second module is still programmed under IDL language. It simulates the whole data processing that has to be carried out in the case of the off-axis TI. First the external PSF file is read, and the OTF is computed through inverse Fourier transform, as shown in Eq. (4). Then the third term in Eq. (5) is isolated and re-centred on the origin of the OXY plane. Phase extraction is achieved by inverting Eq. (9), leading to numerical values of k $\Delta(x,y)$ at every pupil sampling point. The WFE $\Delta(x,y)$ is then easily deduced, and compared with the reference external file generated by the first module. WFE maps can be subtracted one another, displaying "error maps" whose characteristics – in both Peak to Valley (PTV) and Root Mean Square (RMS) – are helpful to estimate and delimit the final measurement accuracy of the method, and pupil areas where the main errors occur.

Further processing allowing to retrieve the individual misalignments of each facet is feasible, although not yet implemented.

3) Similarly, the third module is devoted to phase-shifting TIs. Four external PSF files are read and Fourier transformed, then linearly combined following relation (12). The remaining steps are similar to those of the off-axis TI in the second module.

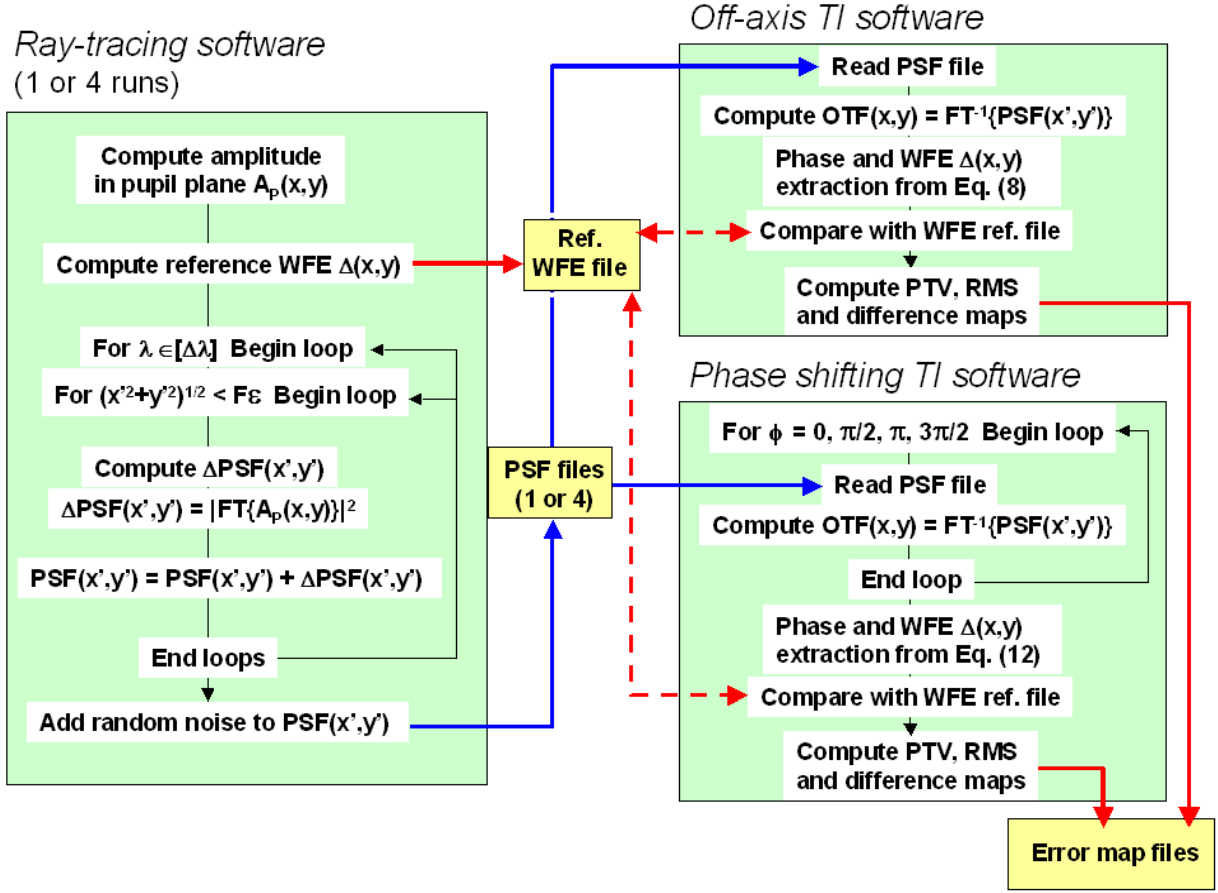

**Figure 4: Block-diagram of TIs computer model**

Having currently developed a specific optical model in order to estimate the systematic errors of both types of TIs, and provided some theoretical relationships dealing with their random errors (see section 3.1), time is coming for numerical applications and discussion. Let us start with five-meters class telescopes, even if they do not represent the future of ground observatories, and try to transform them into TI, of either off-axis or phase-shifting types. The major characteristics and hypotheses are summarized in Table 1 together with some typical parameters of working CCD arrays. A sun-like star is observed in the V-band, which central wavelength is 0.55 µm. The radius of the reference pupil r is matched to the Fried's radius of turbulent cells $r_0$. We choose $r = r_0 = 0.25$ m, which represents excellent conditions of seeing, and ensures a spatial resolution of ten on the main pupil area. The integration time is set to 0.01 second in order to stay compatible with an AO operational system. From Table 1, the major differences between off-axis and phase-shifting versions are the pixels size and sampling of the required detector arrays, and radiometric efficiency of the TI. Here the phase shifting version has the great advantage of requiring no additional optics when the reference pupil can be directly integrated into the primary mirror structure [2]. However this benefit might be somewhat reduced if there is no open access to the telescope image plane for acquiring PSFs. In that case the use of beam-splitting or dichroïc plates becomes mandatory, and radiometric efficiency will decrease accordingly.

Table 1: Typical parameters used for 5-meter class TIs

| Parameter | Off-axis TI | Phase-shift TI | Unit |
|---|---|---|---|
| Central Wavelength | 0.55 | 0.55 | µm |
| Main Telescope Diameter | 5 | 5 | m |
| Reference Pupil Radius | 0.25 | 0.25 | m |
| Integration Time | 0.01 | 0.01 * | sec |
| Radiometric Efficiency | 0.1 | 0.4 | none |
| Pixel Number | 1024 x 1024 | 256 x 256 | |
| Pixel Size | 12 x 12 | 12 x 12 | µm |
| Array Detector Size | 6 x 6 | 1.6 x 1.6 | mm$^2$ |
| Dark Current Noise | 0.1 | 0.1 | elec/phot/sec/pixel |
| Read-Out Noise (RON) | 5 | 5 | elec/phot/pixel |

\* Total time composed of 4 individual acquisitions of 2.5 msec

We assume that the main aperture (i.e. the primary mirror of the telescope) is constituted of six hexagonal segments surrounding either a central, reference mirror in the case of the off-axis TI, or the reference pupil itself on the phase-shifting TI. Each hexagon has 810 mm side and is affected with variable piston errors, which are, like in Refs. [1-2], equal to 0.498 λ, 0.25 λ, -0.25 λ, -0.498 λ, -0.25 λ and 0.25 λ turning clockwise from the Y-axis. Such WFE reference map is illustrated in Figure 5 and our purpose is now to evaluate associated uncertainties and error maps. The first bias is expressed by Eq. (21), and is inherent to the Dirac approximation, generating highest errors near the boundaries of the telescope segments: there the convolution product between the main and reference pupils will average meaningless data found outside of the segments with relevant information located inside. This effect is also encountered near the telescope central obscuration and at the spider supporting legs. But as long as only piston errors are concerned, however, the corrugated boundary areas can easily be removed from the error maps, multiplying them with a pre-computed mask file. Hence the global PTV and RMS numbers are sensibly improved, while the numerical determination of pistons remains quite accurate. An example is shown in Figure 6 for an off-axis TI. The intrinsic error has been sensibly improved with respect to Refs. [1-2]. Actually, the computing size of the hexagon was reduced to 780 mm, which corresponds to a ratio 0.962 along the facet side and to an effective fill factor of 0.925 for reflective facets areas. The previous masking file is conserved for all numerical simulations presented in this section: we assume that induced bias errors on piston values are proportional to (1. – 0.925) Δ(x,y) = 0.075 Δ(x,y) at worst, and can therefore be neglected.

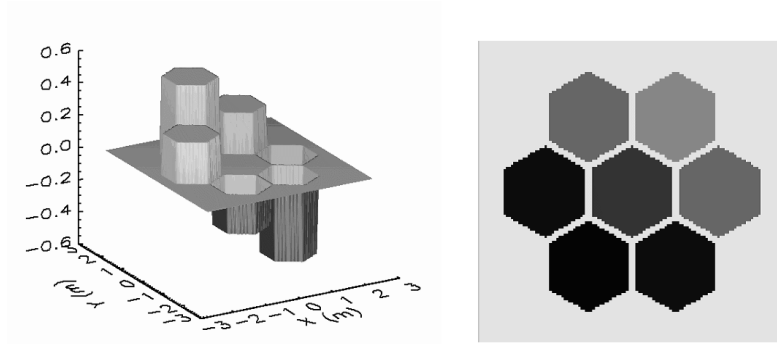

**Figure 5: WFE reference map used to evaluate TI measurement accuracies (PTV = 0.996 $\lambda$, RMS = 0.353 $\lambda$)**

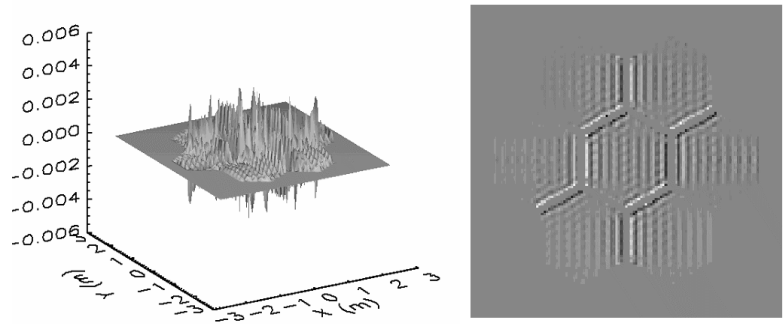

**Figure 6: Intrinsic measurement error of an off-axis TI (PTV = 0.009 $\lambda$, RMS = 0.001 $\lambda$)**

As mentioned in section 2.3, the spectral width of the observed sky-object plays an important role in bias uncertainties since OTFs in Eq. (22) are integrated over a finite bandwidth $\Delta\lambda$, thus averaging and scrambling the side lobes of the acquired PSFs, consequently losing information enclosed in those areas. For both types of TIs, we have performed series of numerical simulations, starting from the monochromatic case, and progressively increasing spectral ranges $\Delta\lambda$ until the method finally fails to retrieve the original reference WFE of Figure 5. The resulting measurement accuracy curves are shown in Figure 7, while Figure 8 illustrates a typical error map obtained with an enlarged spectral bandwidth.

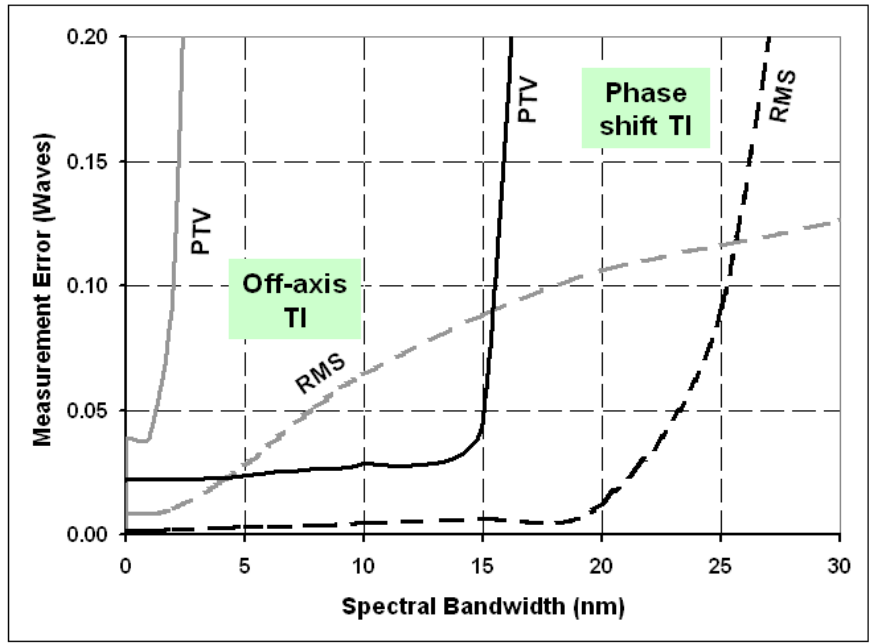

**Figure 7: Measurement accuracy as function of spectral bandwidth $\Delta\lambda$ (black lines: phase-shifting TI; gray lines: off-axis TI; solid lines: PTV values; dashed lines: RMS values)**

The curves in Figure 7 describe dependencies of the bias error term $\delta\Delta(x,y)$ – characterized by its PTV and RMS numbers – with respect to the useful spectral range and the TI family (i.e. off-axis or phase-shifting). A first spectacular conclusion is indeed the apparent superiority of the phase shifting TI with respect to the off-axis version, since the latter is actually limited to extremely narrow spectral bands. In both cases, it can be clearly seen that accuracy slowly deteriorates as bandwidth increases, until a particular threshold of $\Delta\lambda$ is attained. Then PTV curves suddenly reach up aberrant values and RMS numbers follow similar tendencies, while the corresponding error maps reveal strong discontinuities of the retrieved WFEs (not shown herein). The obtained numerical results are helpful to determine acceptable limits on the spectral domain of an observed punctual source of light. Hence we shall define thresholds $\Delta\lambda$ equal to 2 nm and 10 nm respectively for off-axis and phase-shifting TIs, which correspond to effective ratios $\Delta\lambda/\lambda$ of 0.4 % and 2 %. It must be noted that over such limited ranges, spectral bias error easily fulfills classical WFE criteria such as those defined by Rayleigh (PTV < 0.250 $\lambda$) or Maréchal [23] (RMS < 0.075 $\lambda$), stating that an optical system can be considered as diffraction-limited. Therefore the spectral bias is not significant as long as the here above limits are not overcome.

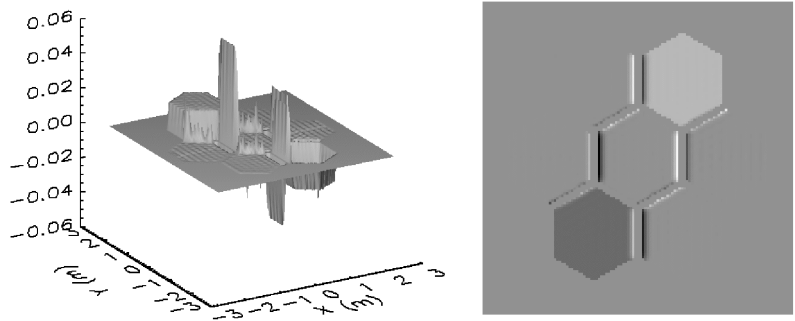

**Figure 8: Typical error map for enlarged spectral bandwidth (Off-axis TI with Δλ = 2 nm, PTV = 0.104 λ and RMS = 0.010 λ)**

The third systematic error cited in section 2.3 is related to the angular size of target star, and therefore to its photometric properties. Here measurement errors $\delta\Delta(x,y)$ are actually governed by two contradictory trends. First the bias term is introduced according to Eq. (23), growing up with the angular radius $\varepsilon$ of the observed star. But increasing the size of the object also improves photon noise, and this effect has to be taken into account using Eqs. (18-20), from which it can be assumed that $\delta\Delta(x,y)$ is inversely proportional to $\varepsilon$ in first approximation. Both bias and noise terms are plotted together in Figure 9 for each type of TI. It can be seen that for a five-meter class telescope operating in AO regime, the best compromise between random and bias uncertainties is attained around 0.75 ± 0.05 mas, which typically corresponds to stars of magnitude around 4 and 5 respectively for off-axis and phase-shifting TIs. As an illustration, Figure 10 depicts a typical error map obtained in the presence of pure photon noise, for $\varepsilon = 0.9$ mas. It must be highlighted that other random components (such as RON and dark current noise) were also introduced into the model without any significant effect on the achieved results and conclusions. In other words, random errors are dominated by photon noise.

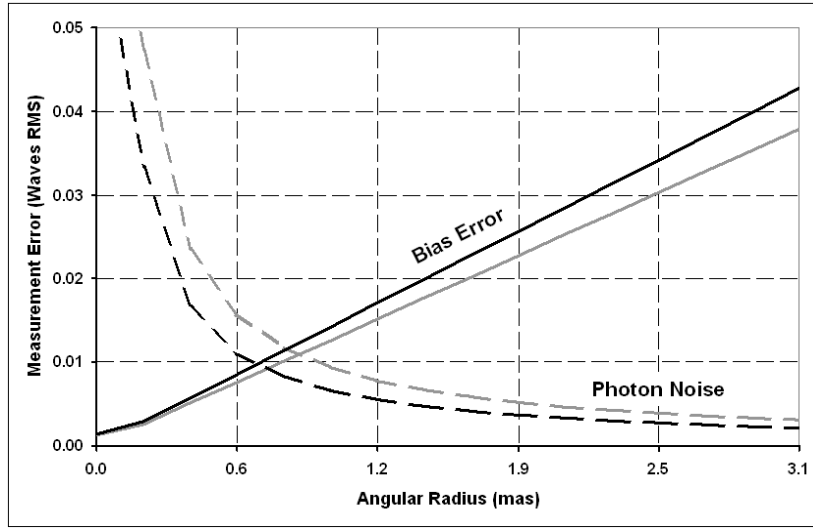

**Figure 9: Measurement accuracy as function of angular radius of the observed sky-object (black lines: phase-shifting TI; gray lines: off-axis TI; solid lines: RMS bias error; dashed lines: photon noise contribution)**

At this point of the study, we have developed different numerical models in order to estimate the performance of a TI. A full analytic approach was followed for estimating random noises (see section 3.1), while systematic errors due to the Dirac approximation, spectral range and geometrical etendue of the stellar beam were evaluated numerically. It was shown that maximal relative spectral widths of 0.4 % and 2 % (depending on the TI family) can be tolerated for the observed sky-object, while a maximal apparent size around 0.75 mas (and thus minimal magnitude around 4-5) is imposed for a 5-m, AO assisted telescope. It can be deduced that, as long as the previous conditions are fulfilled, the effect of systematic errors are negligible when compared to photon noise errors, and that the latter are adequate for defining the operational characteristics of larger TI facilities. This is the goal of the next section.

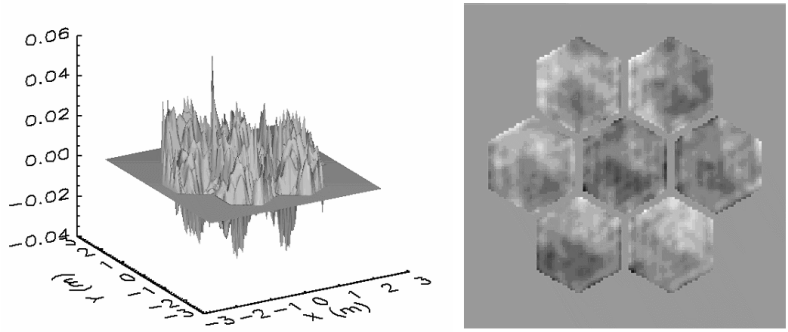

**Figure 10: Typical error map in the presence of photon noise (Angular radius = 0.9 mas, PTV = 0.079 $\lambda$, RMS = 0.009 $\lambda$)**

## 4. More numerical results and discussion

Random noise simulations such as those presented in the previous section proved to be fully relevant in defining acceptable limits on some critical parameters driving a TI performance. Herein are studied successively the influence of the main aperture diameter, magnitude of the observed star, atmospheric turbulence, and total integration time on the detector.

### *4.1 Telescope diameter and limiting magnitude*

In the prospect of building on-ground Extremely Large Telescopes (ELTs), or an imaging hyper-telescope using densification technique, large values of the diameter D = 2 R shall be expected, typically ranging from 30 m to 50 m for the TMT [24] and European ELT [25] projects. Such facilities will need to be equipped with new generation adaptive optics, possibly incorporating high spatial resolution WFS and deformable mirrors with several thousand actuators. In this section are examined the capacities of a TI for such purposes. Using only the random noise numerical model based on Eq. (18) to (20), RMS measurement accuracies for both TI families are plotted in Figure 11 and Figure 12 as a function of the main aperture and reference pupil diameters, for an integration time $\tau$ = 0.01 second. As the RMS error $\delta\Delta(x,y)$ is inversely proportional to the contrast factor C and the latter is equal to the ratio $S_r/S_R$, it can be instantly seen that WFE measurement errors are increasing with $S_R$, i.e. the full optical area of the ELT.

Moreover, since the optical surface $S_r$ of the reference pupil should stay matched to Fried's parameter $r_0$, it follows that larger telescopes will be affected with higher measurement uncertainties if TI principle is applied. Both Figures 11 and 12 clearly illustrate this unfavorable tendency. However, it must be emphasized that:

1) For medium seeing conditions ($r_0$ = 0.1 m), the plotted curves shows that Maréchal's criterion (RMS < 0.075 λ) is fulfilled when the main aperture is equal to 5 m for the off-axis TI, and 9 m for its phase-shifting version. Hence the method would be suitable for the present generation of telescopes, but not for ELTs. These results are consistent with those obtained in section 3.3.
2) For excellent seeing conditions ($r_0$ = 0.25 m), numerical results demonstrate that TI method remain efficient with telescope larger than 30 m in their off-axis configuration, and 50 m for phase-shifting TIs.
3) Also presented is the case of outstanding seeing $r_0$ = 0.5 m, which can probably be only observed on exceptional astronomical sites and in the infrared region of the electromagnetic spectrum. Here it can be thought as an ultimate accuracy limit to be reached, or resulting from a distinct, low-order AO correction system.
4) Figure 12 clearly highlights the advantage of the phase-shift TI with respect to the off-axis version: as a rule of thumb we notice that a 50-meters aperture phase-shifting telescope-interferometer shows performance equivalent to a 30-meters off-axis TI, while a 50-meters off-axis TI would not fulfill Maréchal's criterion.
5) Figure 12 also tell us that for such telescopes, the minimal diameter of the reference pupil is around one meter. For such dimensions, the reference arm of the TI shall be equipped with its own, low-order AO system, in order to minimize differential WFE between main and reference apertures [1].

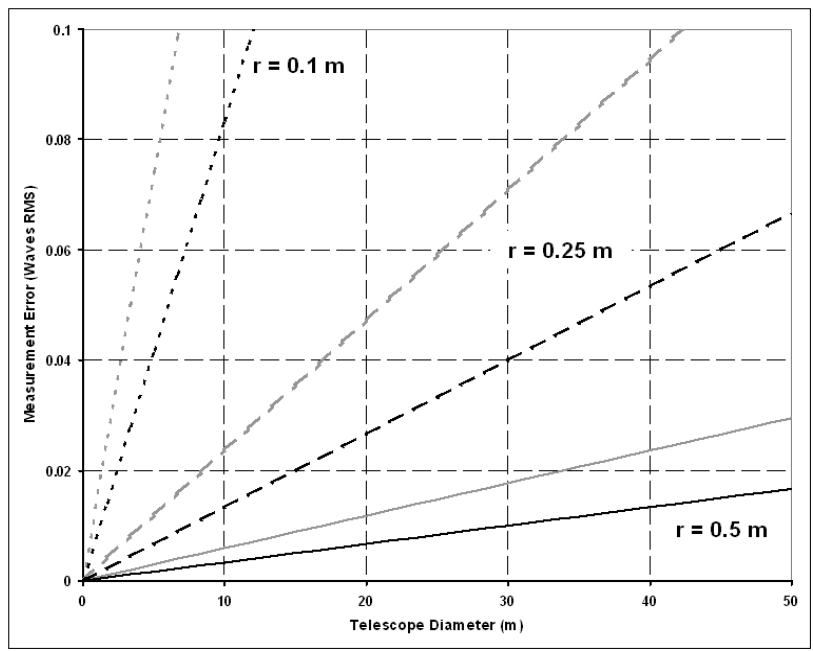

**Figure 11: Measurement accuracy as function of main telescope diameter (black lines: phase-shifting TI; gray lines: off-axis TI; dotted lines: r = 0.1 m; dashed lines: r = 0.25 m; solid lines: r = 0.5 m)**

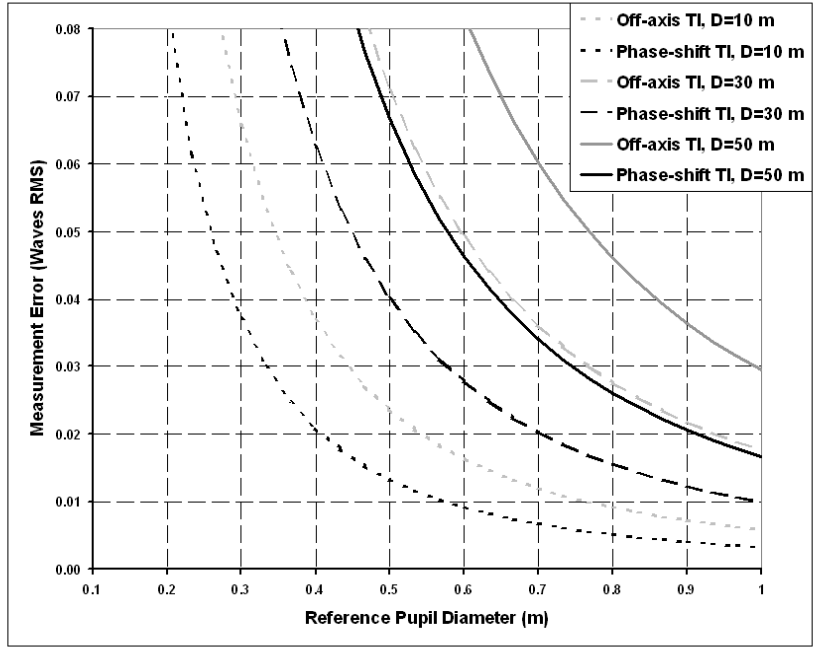

**Figure 12: Measurement accuracy as function of reference pupil diameter (black lines: phase-shifting TI; gray lines: off-axis TI; dotted lines: D = 10 m; dashed lines: D = 30 m; solid lines: D = 50m)**

For a given telescope size, the accuracy of the TI method is also affected by the magnitude of the target star. Here again the main error contribution comes from photon noise, since SNR is often proportional to its square root – at least in first approximation. Let us consider an ELT of diameter D = 30 m and keep integration time to identical value $\tau$ = 0.01 second. Relationships (18) to (20) thus allow to compute the WFE retrieval error in RMS sense, as function of the irradiance of the observed star. The results are shown in Figure 13 for two different reference pupil sizes (r = 0.25 m and r = 0.5 m). The case of medium seeing (r = 0.1 m) is not presented there since it was shown that TI method is restricted to 5-10 m class telescopes in such conditions. It can be seen, however, that for $r_0$ = 0.25 m, Maréchal's criterion is fulfilled for target stars of magnitude 8 and 15, respectively for the off-axis and phase-shifting TIs. Generally speaking, these numbers are not adequate to obtain a good sky coverage, however they might be sufficient for the special purpose of hunting extra-solar planets around nearby stars from a 30-m large, ground-based ELT. In addition we must stress again the superiority of phase-shifting TIs with respect to off-axis TIs for a given set of geometrical parameters, and this assertion will remain valid in section 4.3.

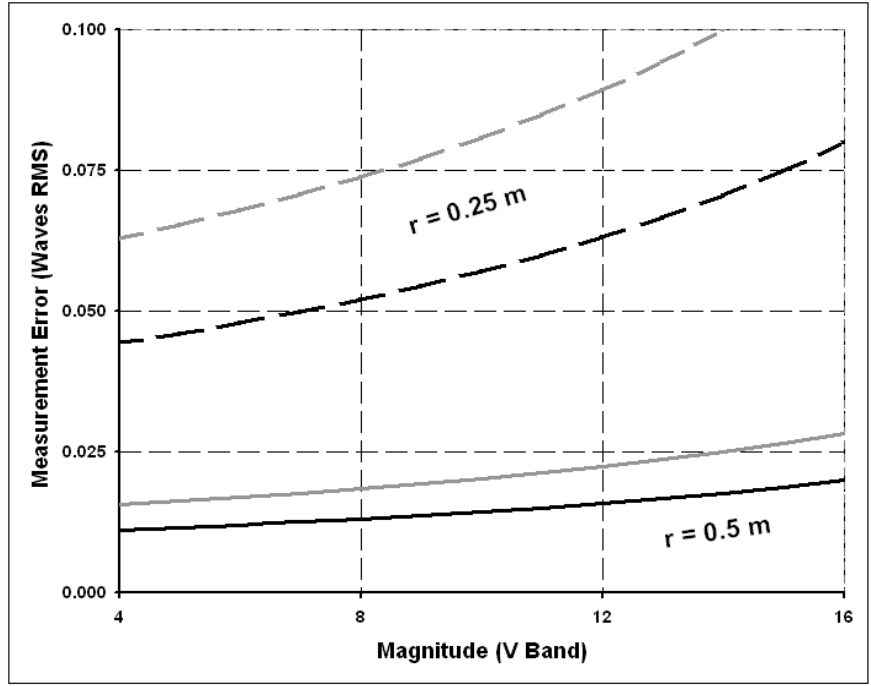

**Figure 13: Measurement accuracy as function of star magnitude (black lines: phase-shifting TI; gray lines: off-axis TI; dashed lines: r = 0.25 m; solid lines: r = 0.5 m)**

## 4.2 Atmospheric seeing as an error source

So far atmospheric turbulence was not considered as a true error item, since in AO operation it usually stands for the major scope of WFE measurements. It must be pointed out, however, that because the spatial resolution of TIs is limited by the size of the reference pupil, on one hand, and photon noise increases as reference aperture gets smaller, on the other hand, TI method is probably not suitable for very high order AO applications – and such drawback will not disappear when an AO system is added to the reference pupil arm. Another point of concern is the coupling of atmospheric seeing with cophasing errors of the segmented telescope, and the way they can be distinguished. Finally, we are also interested in enlarging the sky coverage – already discussed in previous section – by extending global exposure times. Therefore TI principle shall not only be envisaged for AO operation, but might also be thought as an active optics method for cophasing ELTs, such as those described in Refs. [11-15]. To provide answers to these questions, additional simulations including seeing perturbations were performed and are summarized below.

Let us consider an off-axis TI affected with the same cophasing errors than in section 3, and the following basic parameters for atmospheric turbulence: Fried's radius $r_0$ is equal to 0.1 meter with a wind speed of 10 m/s and an outer scale of 20 m. Acquisition time of each single PSF stays unchanged (10 milliseconds), while total integration time is 6 seconds, which corresponds to 600 consecutive WFE measurements. For the sake of illustration, computations were also carried out with other values of $r_0$ (0.25 and 0.5 meters), confirming the forthcoming conclusions. The latter are summarized in Figure 14: on top is sketched a single WFE acquisition, before (left) and after phase unwrapping with a classical algorithm (right). Piston errors remain clearly observable at the edges of each telescope segment and can therefore be estimated. The obtained numerical results for long exposure times are displayed at the bottom row of Figure 14, for two different data processing modes:
- In first approach, 600 PSFs are averaged before OTF calculation, leading to a complete loss of phase information (bottom left). The reason is that PSF modulation is actually scrambled by numerous disturbed phase screen acquisitions.
- The second computation mode is somewhat inspired from the technique of speckle-interferometry: it consists in averaging unwrapped phases of the different OTF maps, each being calculated from one single PSF at a 10 ms acquisition rate. It is expected that seeing perturbations are averaged and vanish over the full sequence of 600 measurements, then revealing only telescope cophasing errors. However, the example presented at the bottom right of Figure 14 shows that this is untrue, since the final phase map does not reproduce initial piston errors, on one hand, and various tilt and focus residuals can be observed on each reflective facet, on the other hand. In fact, it seems that piston errors were washed away by atmospheric seeing.

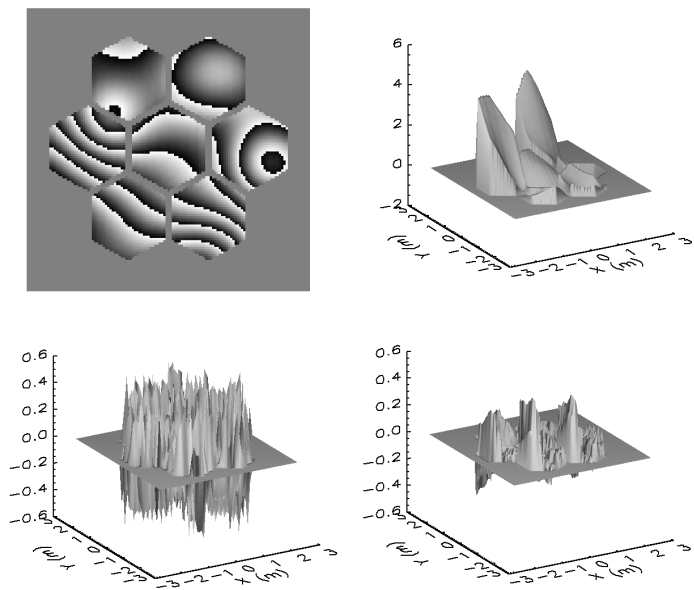

**Figure 14: Simulation of WFE retrieval on long exposure times. Top: instantaneous WFE acquisition with Fried's radius $r_0$ = 500 mm, PTV = 6.124 $\lambda$, RMS = 1.230 $\lambda$ (left and right, before and after phase unwrapping). Bottom: Piston retrieval errors with $r_0$ = 100 mm (left and right, averaging on PSFs – PTV = 1.000 $\lambda$, RMS = 0.265 $\lambda$ – and on unwrapped OTF phases – PTV = 0.968 $\lambda$, RMS = 0.114 $\lambda$). Both latter maps show that initial cophasing errors are not retrieved**

Obviously these results need to be confirmed, since possible errors may originate from a too short PSFs measuring sequence, or inadequate phase unwrapping algorithm. However they cast doubt on the real capability of TIs to perform long-exposure WFE measurements on a ground-based ELT. Hence it shall be assumed – at least provisionally – that TIs are not recommended for active optics applications, and the best operating parameters remain those defined in section 4.1.

## *4.3 Total integration time*

Let us disregard ground applications for a short section, and consider the case of a space telescope. Then PSFs measured by the detector array are no longer perturbed by the atmosphere, and the total integration period can be extended to a few minutes of telescope observing time. We assume that a space telescope of diameter D = 10 m is equipped with an extra-solar planet finding instrument (e.g. a coronagraph). Then TI measurement technique – whatever its type, although the off-axis version might be preferred because it exhibits no central obstruction – can be used to sample the pupil with several hundred of points, consequently improving WFE spatial resolution. This can be achieved by simply decreasing the size of the reference pupil with respect to the main aperture, as shown in Figure 15. We can see for example that spatial resolutions around 1000 x 1000 can be obtained on both types of TIs over integration times around 10 minutes, while 400 x 400 WFE sampling points are acquired in a few seconds only. Hence the technique seem to be well suited for regular calibration and co-phasing of the segmented mirrors constituting a space-borne coronagraph or hyper-telescope.

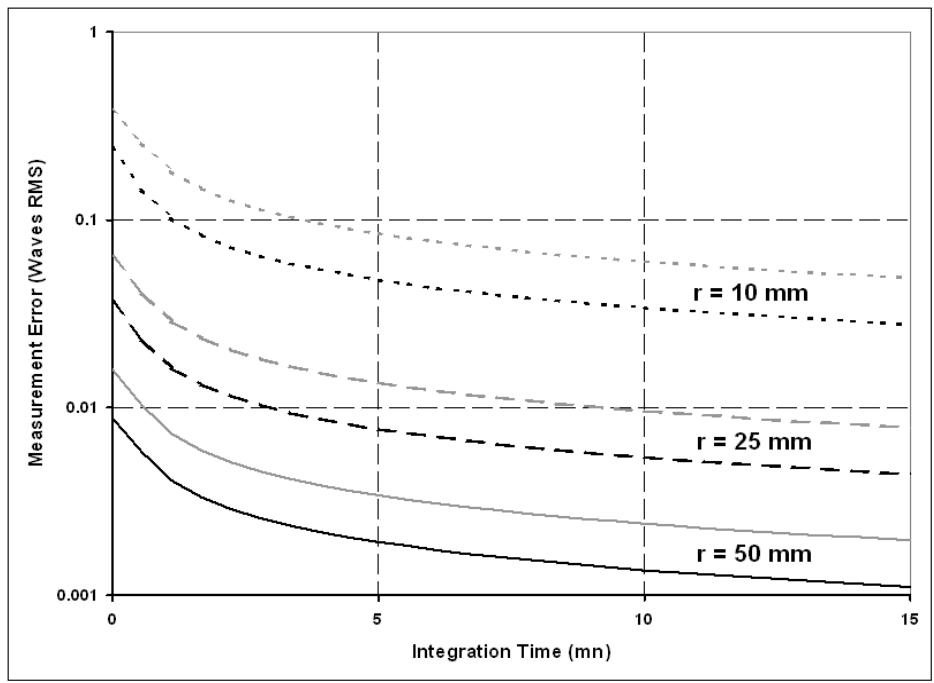

**Figure 15: Measurement accuracy as function of integration time (black lines: phase-shifting TI; gray lines: off-axis TI; dotted lines: r = 10 mm; dashed lines: r = 25 mm; solid lines: r = 50 mm)**

# 5. Conclusion

Coming after Refs. [1-2], this paper aimed at concluding a trilogy dealing with the principles, performance and limitations of what I named "Telescope-Interferometers" in a previous publication. The basic idea consists in transforming a telescope into a WFE sensing device by giving to sky photons an additional access to the focal plane of the telescope through its exit pupil. This can be achieved in two different ways, namely the off axis and phase-shifting TIs. In the first case a smaller and decentred reference telescope is added aside the main pupil, creating a weakly modulated fringe pattern in the image plane. In the second configuration different calibrated phase shifts are applied over the reference pupil area. In both cases the PSFs measured in the focal plane of the telescope carry information about the transmitted WFE, which is retrieved by fast and simple algorithms suitable to an adaptive optics operational mode.

Herein was evaluated the accuracy of both types of TIs, in terms of noise and systematic errors. Numerical models were developed in order to calculate the influence of driving parameters such as useful spectral range and angular size of the observed star, or noises of the CCD detector array. Some lessons learned from numerical results contradict a few previous statements from Refs [1-2]. We found that RMS measurement error is proportional to the geometrical area of the main telescope, and inversely proportional to detector SNR and reference pupil area. Spectral bandwidth $\Delta\lambda$ and radiating properties of the target star were defined (e.g. $\Delta\lambda < 2$ nm and magnitude comprised between 5 and 15 for a phase-shifting TI in the V band). It was shown that WFE measurement errors are particularly sensitive to photon noise, which rapidly governs the achieved accuracy for telescope diameters higher than 10 m. We studied a few practical examples, showing that TI method is applicable to adaptive optics systems on telescope diameters ranging from 10 to 50 m, depending on the seeing conditions and magnitude of the observed stars. We also discussed the case of a space-borne coronagraph where the TI technique can provide very high sampling of the input WFE within a few minutes of integration time. Finally, as a general rule, phase-shifting TIs were found clearly superior to their off-axis version for what concerns the usable spectral range, maximal magnitude of the target star, achievable accuracies, and the required number of CCD pixels – assuming the same geometrical characteristics.

Other attractive theoretical studies remain to be undertaken around TI matter, such as comparison with different types of WFS – and particularly those depicted in Refs. [3-4] – application to scintillation measurements, or improvements of the method by implementing a deconvolution process in order to get rid of the "Dirac approximation". Among these topics, priority should be given to a full comparative analysis with other WFS in AO or non-AO regimes, involving numerous criteria such as intrinsic measurement errors and noises, spatial sampling and resolutions, operational ranges in terms of seeing conditions and magnitude of the observed sky-object, eventual use of laser guide stars, or phase sensing ability on sparse apertures. This seems to be a mandatory step before TIs prototyping can be envisaged on a dedicated optical testbench, or an existing telescope facility.